\documentclass[twocolumn,trackchanges,twocolappendix]{aastex701}

\begin{document}

\title{Nearby M\&M Binaries Uncovered with the Gaia Multipeak Fraction and Speckle Interferometry}

\author[0000-0001-5038-0089]{Dmitry Chulkov}
\affiliation{Sternberg Astronomical Institute, Lomonosov Moscow State University\\ 119992 Universitetskii prospekt 13, Moscow, Russia,
\href{mailto:chulkovd@gmail.com}{chulkovd@gmail.com}} %
\email[]{chulkovd@gmail.com}

\author[0000-0003-0647-6133]{Ivan A. Strakhov}
\affiliation{Sternberg Astronomical Institute, Lomonosov Moscow State University\\ 119992 Universitetskii prospekt 13, Moscow, Russia, 
\href{mailto:chulkovd@gmail.com}{chulkovd@gmail.com}}%
\email[]{strakhov.ia15@physics.msu.ru}

\author[0000-0003-1713-3208]{Boris Safonov}
\affiliation{Sternberg Astronomical Institute, Lomonosov Moscow State University\\ 119992 Universitetskii prospekt 13, Moscow, Russia, 
\href{mailto:chulkovd@gmail.com}{chulkovd@gmail.com}}
\email[]{\\safonov@sai.msu.ru}

\author[0000-0002-4973-270X]{Vladimir V. Dyachenko}
\affiliation{Special Astrophysical Observatory, 369167 Nizhnij Arkhyz, Russia, 
\href{mailto:dyachenko@sao.ru}{dyachenko@sao.ru}}
\email[]{\\dyachenko@sao.ru}

\author[0009-0009-5465-3182]{Margarita D. Butorina}
\affiliation{Special Astrophysical Observatory, 369167 Nizhnij Arkhyz, Russia,
\href{mailto:dyachenko@sao.ru}{dyachenko@sao.ru}}
\email[]{\\tauri27@yandex.ru}

\author[0000-0002-7790-5022]{Anatoly V. Beskakotov}
\affiliation{Special Astrophysical Observatory, 369167 Nizhnij Arkhyz, Russia,
\href{mailto:dyachenko@sao.ru}{dyachenko@sao.ru}}
\email[]{\\beskakotov@sao.ru}

\begin{abstract}
Multipeak fraction is a strong indicator of stellar multiplicity with subarcsecond angular separation in the Gaia DR3 catalog. We select 18 sources with multipeak fraction above 5\% among stars possibly in the 20 pc solar neighborhood with as yet unconfirmed binarity. We successfully resolve eleven systems of M-type dwarf components in speckle interferometric observations with a 2.5 m telescope at the Caucasian Mountain Observatory of Sternberg Astronomical Institute, Moscow State University. The tentatively estimated companion masses lie in the 0.09 -- 0.23 $M_\odot$ range. All six sources with a multipeak fraction of 6--8\% are unresolved, suggesting an observational bias that affects nearby stars.
\end{abstract}



\section{Introduction} 
Stars in the immediate solar neighborhood are of special interest to astronomers as they allow the most comprehensive, detailed and accurate measurements. This makes them promising targets for the exploration of the exoplanet population, including a search for potential biosignatures, justifying the need for full characterization of these objects. High completeness down to the brown dwarf regime allows an elaborate study of the initial mass function \citep{2024ApJS..271...55K}. However, despite decades of dedicated efforts, the inventory of nearby stars still remains incomplete. In this paper, we address unrecognized multiplicity, one of the existing issues leading to sample incompleteness \citep{2025A&A...693A.228C}. Even the completeness of the 10 pc sample is questionable \citep{2024RNAAS...8..162S}, and the presence of a few missing stellar components within it remains possible \citep{2026MNRAS.549ag838G}.

Multiplicity rates progressively increase with stellar mass \citep{2023ASPC..534..275O}. Even in the low-mass range occupied by the M-type dwarfs \citep{2024ARA&A..62..593H}, around a quarter of stars have at least one companion \citep{2026AJ....172...40C,2026ApJ..1004..224W}. The stellar population is very diverse, which requires a complex approach to multiplicity discovery and characterization \citep{2025A&A...693A.228C}. 

This study focuses on the search for unknown binary stars with the speckle interferometric method. High-angular-resolution observations conducted with speckle imaging are among the principal methods for stellar multiplicity detection and characterization \citep{2021FrASS...8..138S}. This technique is routinely applied to the objects of interest, including exoplanet host stars \citep{2025AJ....169...76M} and prospective Habitable Worlds Observatory targets \citep{2026AJ....171..174H}. Recent studies probing the multiplicity of solar neighborhood stars include the RECONS \citep{2020AJ....160..215V} and POKEMON \citep{2022AJ....164...33C} programs.

In Section \ref{sample}, we start by identifying sources with unconfirmed multiplicity, which are promising targets for high-angular-resolution observations. We describe our instruments, observations, and data reduction in Section~\ref{observations}. The procedure of tentative mass estimation is explained in Section \ref{masses}. The results for unresolved and resolved stars are discussed in Sections \ref{Unresolved} and \ref{Resolved}. Section~\ref{summary} provides a brief summary of our work. The paper concludes with the Appendices, which contain a study of sources with elevated multipeak fraction in Gaia DR3 (Section~\ref{allsky}), record the observations of reference stars (Section~\ref{app:reference}), and provide a comparison of theoretical isochrones with empirical data for M dwarfs (Section~\ref{isochrones}). 

\section{Observational targets}
\label{sample}
We attempt to select the most promising targets among stars with as yet unconfirmed multiplicity. Following the survey of the Pleiades members \citep{2025AJ....169..145C}, we noticed that the multipeak fraction in Gaia~DR3 is a powerful indicator of subarcsecond multiplicity (Section \ref{Multipeak}). This inspired us to verify whether we can directly confirm the binarity of sources with an elevated multipeak fraction in the solar neighborhood.

\subsection{Volume-limited sample}
A number of studies, dating back at least to the \cite{1957MiABA...8....1G} catalog, have attempted an inventory of the solar neighborhood and the construction of volume-limited samples. Although the results of the Gaia mission \citep{2016A&A...595A...1G} drastically improved the census of nearby stars \citep{2021A&A...649A...6G}, this remains a non-trivial task that requires dealing with inevitable errors in parallax measurements \citep{2015PASP..127..994B}. In this study, we largely put aside the analysis of stellar distances and rely on the Fifth Catalogue of Nearby Stars (CNS5) compiled by \cite{2023A&A...670A..19G}. This catalog aims to include all stars and brown dwarfs within 25 pc of the Sun. 

Trigonometric parallaxes from Gaia early Data Release 3 (eDR3) \citep{2021A&A...649A...1G} remained the most up-to-date and precise data available for most nearby stars while this paper was in preparation. A zero-point offset for Gaia parallaxes and an inflation factor for their uncertainties are adopted in CNS5 to improve sample selection; therefore, the values may differ slightly from the original Gaia data. CNS5 relies on Hipparcos \citep{1997A&A...323L..49P} and ground-based observational data for stars with unreliable or unavailable astrometric solutions in Gaia. This approach is essential for binaries, as their astrometric solutions are often compromised \citep{2022MNRAS.513.2437P}.

\subsection{Multipeak fraction}
\label{Multipeak}
During its mission, the Gaia satellite continuously surveyed the celestial sphere in scanning mode. Due to the different angular resolution along the scan and in the orthogonal direction, the detectability of a double star during a field-of-view passage depends on the scanning direction and the orientation of the system \citep{2023A&A...674A..25H}. The multipeak fraction \text{\it ipd\_frac\_multi\_peak} ($f_{\rm MP}$) indicates the percentage of observations for which the image parameter determination algorithm  identified a double peak for a given source \citep{2021A&A...649A...2L}. By definition, $0 \leq f_{\rm MP} \leq 100\%$, and $f_{\rm MP}=0$ for the vast majority of sources in Gaia DR3. \cite{2022MNRAS.513.5270P} pointed out that this parameter should be mostly sensitive to binary stars with angular separations of $0.18\arcsec \lesssim \rho \lesssim 1\arcsec$: systems with smaller separations remain unresolved regardless of the scan direction, while components of wider pairs receive separate solutions. 

\cite{2023AJ....165..180T} recognized $f_{\rm MP}$ as a powerful diagnostic for existence of close companion, which nevertheless receives limited attention in the literature relative to RUWE, a well-known binarity indicator (Section \ref{ruwe}). \cite{2024OJAp....7E.100E} confirmed that most sources associated with $\rho< 0.2$\arcsec \ systems have $f_{\rm MP}=0$ in Gaia~DR3. Consequently, sources with an elevated multipeak fraction should be confidently resolvable with our 2.5 m telescope (Section \ref{observations}). Previously, in the Pleiades, we successfully confirmed binarity for all sources with $f_{\rm MP} >4\%$ \citep{2025AJ....169..145C}.

However, an elevated multipeak fraction does not guarantee physical binarity, as it is also sensitive to crowding. \cite{2023AJ....166..218M} have shown that $f_{\rm MP}$ is a less effective indicator of genuine binarity near the Galactic plane where the stellar density is the highest. According to Table 5 in \cite{2024AJ....167...56C}, at least nine stars remained unresolved with the 4.3 m telescope despite $f_{\rm MP} >5\%$. Hence, the follow-up observations are pivotal to confirm physical binarity.

\subsection{Other binarity indicators}
\label{ruwe}
While we pay most of our attention to the multipeak fraction in this study, there are other, no less important, indicators of binarity. All astrometric data published in the Gaia DR3 main source catalog assume single-star motion. The two-parameter solutions without a parallax estimate were used as a fallback option when the initially attempted five- and six-parameter solutions converged poorly, which often happens for binary stars. Indeed, having a two-parameter solution strongly correlates with $f_{\rm MP}$: while the mean multipeak fraction is within 2\% for $G<15$~mag stars with five-parameter solutions, this value increases to $40 \lesssim f_{\rm MP} \lesssim 60$\% for two-parameter solutions (Tables 4 and 6 in \cite{2021A&A...649A...2L}). 

Another metric commonly used as an indicator of binarity \citep{2020MNRAS.496.1922B} is RUWE (renormalised unit weight error), which efficiently quantifies the difference between the motion of the photocenter and the center of mass for a stellar system \citep{2020MNRAS.495..321P}. A RUWE $>$ 1.4 threshold is often adopted to select solutions with significant deviation from the single-star model \citep{2021ApJ...907L..33S}. RUWE is particularly sensitive to systems with an orbital period close to the total data collection time of 34 months for Gaia DR3. These binaries usually have an angular separation too small to trigger an elevated~$f_{\rm MP}$.

\subsection{Sample definition}
\label{criteria}
The target list for the observational program was built to meet the requirements described below. We admit that the conditions are largely motivated by the desired size of the resulting sample and are somewhat arbitrary. 

\begin{itemize}

\item The star enters the Fifth Catalogue of Nearby Stars \citep{2023A&A...670A..19G}; the upper limit of the three-standard-deviation range for its parallax exceeds 50 mas. Inadvertently, we also included CNS5 1692, which in fact barely misses this threshold (Section \ref{CNS5_1692}). We admit some selected sources likely lie beyond the 20~pc distance from the Sun.  

\item The star is not marked as multiple in the CNS5. A further search in the literature was made to ensure the lack of confirmed companions. In particular, binary systems directly resolved by Gaia or included in the Washington Double Star catalog \citep{2001AJ....122.3466M} are omitted. As an exception, we observe CNS5 2081, despite its presence in the WDS catalog (Section \ref{CNS5_2081}).

\item The multipeak fraction in Gaia DR3 is $f_{\rm MP} >5\%$.

\item The source has an apparent magnitude of $G<16$. This condition is motivated by the faint end limitation of our observational technique (Section~\ref{observations}).

\item Declination above $-15^\circ$. This is determined by the telescope location at $43^\circ44^\prime$ northern latitude.
\end{itemize}

\begin{deluxetable*}{lccccccccc}
\tablecaption{Sample sources sorted by parallax \label{tab:targets}}
\tablewidth{0pt}
\tablehead{
\colhead{2MASS} & \colhead{CNS5} & \colhead{Gaia DR3} & \colhead{$\varpi$} & \colhead{$\sigma_\varpi$} & \colhead{IPDfmp} & \colhead{RUWE} &$G^{\rm app}$&$K_s^{\rm app }$&Resolved \\
\colhead{} & \colhead{} & \colhead{} & \colhead{(mas)} & \colhead{(mas)} & \colhead{\%} & \colhead{} &\colhead{(mag)}&\colhead{(mag)}&
}
\decimalcolnumbers
\startdata
23415498+4410407&5844&1926461164913660160&316.54&0.05&8&1.027&10.38&5.93&0\\
21462206+3813047&5362&1952802469918554368&141.94&0.02&8&1.193&10.83&7.20&0\\
22094029--0438267&5478&2627117287488522240&113.49&0.03&6&1.354&9.24&5.59&0\\
08255285+6902016&2081&1097353325107339776&81.37&0.44&53&4.628&13.75&9.16&1\\
04520573+0628356&1218&3288082758293022848&80.61&0.03&6&1.165&10.71&6.94&0\\
00304867+7742338&142&564184743637112832&76.4&4.0&27&--&14.51&9.56&1\\
05053461+4648017&1280&207033680124098048&56.84&0.63&69&4.823&15.90&10.2&1\\
07193126+3249482&1816&893214796542513280&54.91&0.03&18&1.825&9.36&6.38&1\\
14020606+2102454&3470&1247332788512753536&54.15&0.04&15&1.685&11.97&8.08&1\\
07075044+6712046&1776&1102868303634261632&52.76&0.12&8&6.819&10.29&7.03&0\\
18562628+4622532&4709&2107257297843283584&51.19&0.03&44&1.833&12.61&8.72&1\\
12183488--0625232&3021&3584226658929871232&51.1&0.05&7&1.292&12.41&8.71&0\\
18341610--1426154&4593&4103719347391021952&50.7&0.09&35&0.833&13.56&9.61&0\\
18532537+0250490&4695&4279261876207080576&50.26&0.32&36&12.457&11.44&8.1&1\\
12531943--0519523&3180&3678456111101341056&49.5&8.9&100&--&12.41&8.12&1\\
09084605+6635384&2267&1068825362052486784&49.33&0.64&10&38.98&11.87&8.41&1\\
07420968+5004275&1907&982051285912835200&48.61&0.49&58&20.205&11.18&7.62&1\\
06481555+0326243&1692&3127357248060906368&40.2&3.2&40&--&15.54&10.59&1\\
\enddata
\tablecomments{Columns: (1--3) Identifiers in 2MASS, CNS5 and Gaia DR3 catalogs, respectively; (4--5) Parallax and its uncertainty according to CNS5; (6--8) Multipeak fraction, RUWE (not provided for sources with two-parameter solutions) and apparent $G$ magnitude from Gaia DR3; (9) Apparent $K_s$ magnitude from 2MASS. Column (10) indicates whether the source was successfully resolved during speckle observations. The selection criteria are outlined in Section \ref{criteria}.}
\end{deluxetable*}

The resulting list includes 18 entries (Table~\ref{tab:targets}). The selected sources belong to the M~dwarf population, whose masses range from 0.61 down to 0.075 $M_\odot$, although we do not impose a spectral-type cutoff or put any restrictions at the bright end. The M dwarfs roughly represent 3/4 of the entire stellar population in the 10 pc neighborhood \citep{2024ARA&A..62..593H} and are considerably less studied than FGK-type stars; many of the red dwarfs lack prior high angular resolution observations.

\section{Observations}

Below we describe the dedicated observations made with the SAI-2.5 (Section \ref{sai25}) and BTA-6 (Section \ref{bta}) telescopes. Furthermore, for two targets observational data were found in the literature (Sections \ref{CNS5_1907} and \ref{CNS5_2081}).

\label{observations}
\subsection{Observations with SAI-2.5}
\label{sai25}

Speckle interferometric observations were obtained from July 2025 to October 2026 with the speckle polarimeter installed on the 2.5 m telescope \citep{2020gbar.conf..127S} at the Caucasian Observatory of Sternberg Astronomical Institute, Moscow State University (SAI MSU). Imaging was performed 
with a fast low--noise CMOS Hamamatsu ORCA–Quest detector \citep{2023AstBu..78..234S}; the atmospheric dispersion compensator was employed in all observations. The median seeing, estimated from the FWHM of the average images, was 0.91\arcsec. For $G<12.5^{{\rm mag}}$ sources, 4000 frames with a 30 ms exposure were captured. For the fainter objects, 2000 frames with a 60 ms exposure were acquired. The $512 \times 256$~px frame size and the $20.56$~mas~px$^{-1}$ scale yield a $10.5 \times 5.3$\arcsec \ field of view. Most observations were obtained with an $I_c$-band filter of the Johnson-Cousins photometric system. For brighter stars with $G<9.5$ mag, we used an 880~nm filter with a 70~nm half-width. 

\begin{figure*}[ht!]
\epsscale{0.85}
\plotone{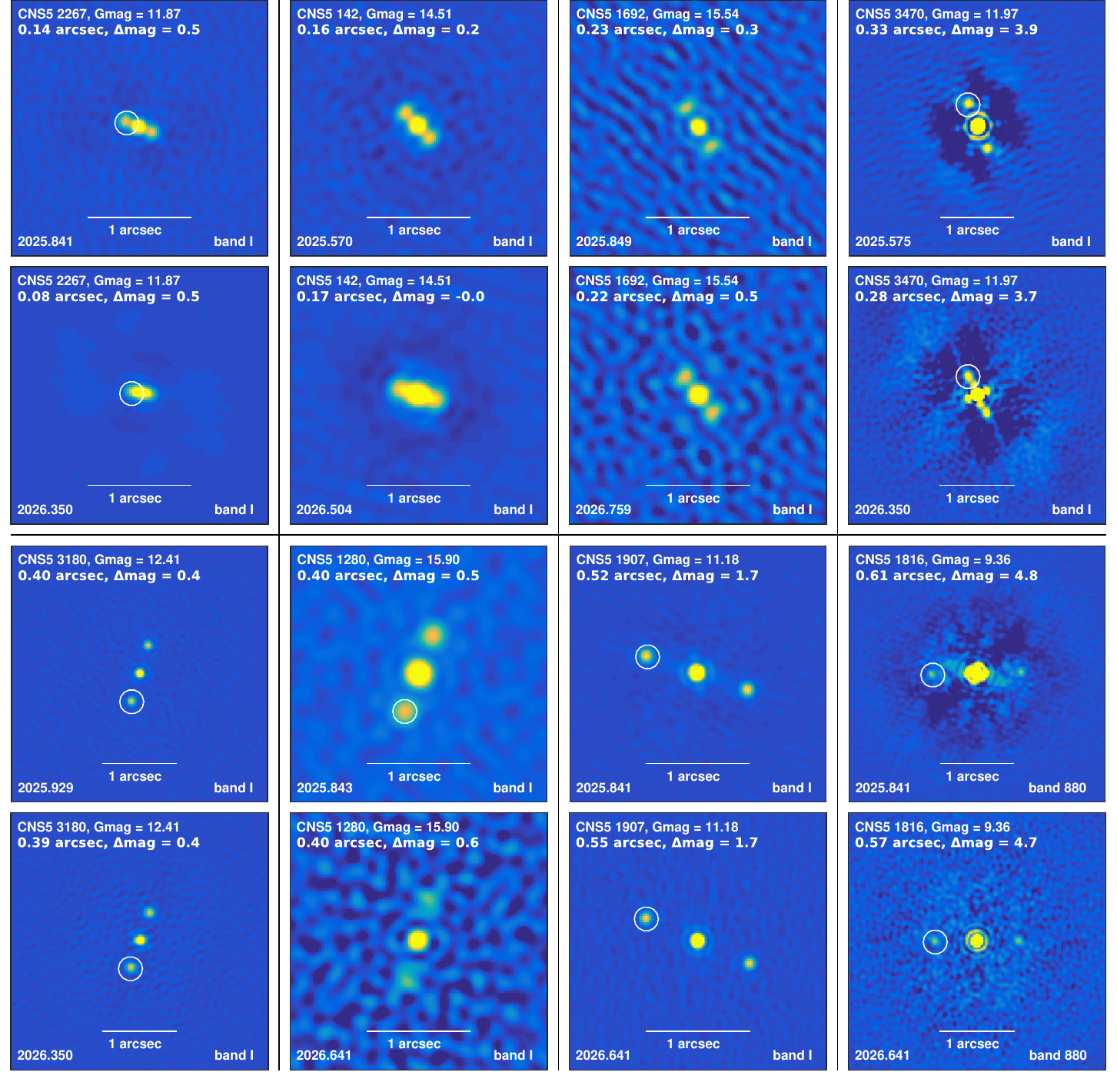}
\caption{Autocorrelation functions computed from SAI-2.5 speckle interferometric observations for resolved binaries. Sources are sorted by measured angular separation, with two epochs shown for each object. A white circle indicates the position of the companion when the 180$^\circ$ ambiguity is successfully resolved. Images of CNS5 4695 and CNS5 4709 are shown in Figure \ref{fig:saibta}.
\label{fig:iconostasis}}
\end{figure*}

The observational technique and data reduction largely follow the routine described in Section 6 of \cite{2025AJ....169..145C}; the resulting autocorrelation functions for our targets are shown in Figure \ref{fig:iconostasis}. Speckle-interferometric imaging enables us to obtain the principal binarity parameters of double stars: angular separation ($\rho$), position angle ($\theta$) and magnitude difference in a given passband (e.g., $\Delta I_c$). Due to the inherent loss of phase information, the position angle is determined with a 180$^\circ$ ambiguity. To resolve this uncertainty, we recover the phase of the object visibility from the average bispectrum using a recursive algorithm \citep{1983ApOpt..22.4028L,2004PASP..116...65T}. This method performs poorly for very close or twin-like pairs with $\rho \lesssim 0.1^{\prime\prime}$ or $\Delta I_c \lesssim 1^{\rm mag}$. Hence, for CNS5~142 and CNS5~1692 the 180$^\circ$ ambiguity is not eliminated. Finally, the detection limits $\varepsilon_{lim}(\rho)$ are calculated as the $5\sigma$ level above the average autocorrelation function value in a ring of radius $\rho$. Any companion with a smaller magnitude difference would be reliably detected.

The reference single stars are used for the normalization of the power spectrum and the estimation of the squared modulus of the object visibility. They are selected within a few degrees of the object to allow a better calibration of atmospheric and instrumental aberrations and observed immediately before and after each target. Reference stars should preferably be one to two magnitudes brighter than the main target and show no indications of binarity in Gaia DR3, with RUWE $<1.25$ and $f_{\rm MP}=0$. Nevertheless, one of the selected sources, 2MASS J07433735+5004098, was serendipitously resolved into a double star with $\rho \sim 0.9\arcsec$. We propose the designation WDS 07436+5004 and provide its binarity parameters in Table \ref{tab:double_star}, along with the detection limits for other reference stars in Table \ref{tab:reference} as part of Appendix \ref{app:reference}. Due to the magnitude difference of $\Delta I_c\sim 5.8$ mag, the faint companion does not compromise the role of this source as a reference for CNS5~1907. 

The nominal uncertainties of the binarity parameters are obtained from a model fit of bootstrapped realizations of the object's estimated visibility squared modulus. The measurements of the magnitude difference between the components are particularly prone to errors. Below we consider them in greater detail as this is essential for the photometric estimation of the mass ratio.  For each individual observation, we obtain an estimate of the magnitude difference along with its nominal uncertainty: $\varepsilon_i \pm \sigma_i$. However, the median error of $\sigma_i=0.04$~mag derived from the power spectra analysis is certainly underestimated. Indeed, the pooled standard deviation for resolved binary stars with repeated ($k=$ 2 or 3) observations in the same filter is twice as large:

\begin{equation}
\sigma_p = \sqrt{\frac{\sum_{i=1}^{k} (n_i - 1)\, \sigma_i^{2}}{\sum_{i=1}^{k} (n_i - 1)}}=0.08 \rm{\ mag} 
\end{equation}

Photometric errors inevitably reduce the accuracy of mass ratio estimates. 
To estimate $q$, we consider observations in the $I_c$ band. For each system, we calculate an inverse-variance weighted mean value and error of magnitude difference according to nominal uncertainties:

\begin{equation}
\begin{array}{l}
\displaystyle \varepsilon =
\frac{\sum_{i=1}^{k} w_i \varepsilon_i}{\sum_{i=1}^{k} w_i},
\qquad w_i = \frac{1}{\sigma_i^{2}}, \\[10pt]
\displaystyle \sigma_n =
\frac{1}{\sqrt{\sum_{i=1}^{k} w_i}}
\end{array}
\end{equation}

This approach allows us to discard the less reliable measurements obtained under unfavorable conditions. The largest uncertainties based on nominal errors belong to CNS5 142 and CNS5 1692, with $\sigma_n \sim 0.10$ mag. Further, following the approach in our Pleiades survey (Section 6.2 in \cite{2025AJ....169..145C}), we cautiously add a systematic uncertainty of $\sigma_s = 0.25$ mag, which therefore dominates the total adopted error: $\sigma=\sqrt{\sigma_n^2+\sigma_s^2}$.

\subsection{Observations with BTA-6}
\label{bta}

\begin{figure}[ht!]
\plotone{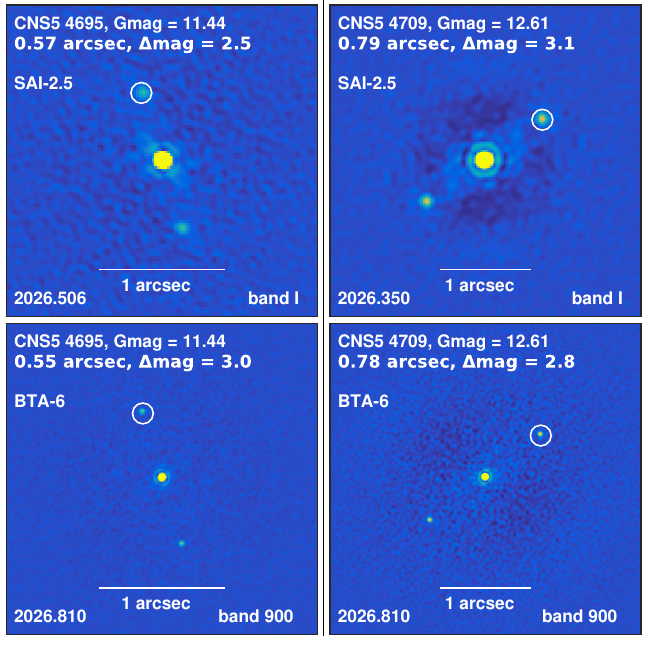}
\caption{Autocorrelation functions computed from speckle interferometric observations of binary stars CNS5 4695 and CNS5 4709 obtained at SAI-2.5 (Section \ref{sai25}) and BTA-6 (Section \ref{bta}). Neither companion was detected at BTA-6 with the 550/50 nm filter. A white circle indicates the companion position derived from the bispectral phase recovery. 
\label{fig:saibta}}
\end{figure}

Two targets, CNS5 4695 and CNS5~4709, were also observed with the 6 m telescope of the Special Astrophysical Observatory of the Russian Academy of Sciences (SAO RAS) on the night of October 23, 2026, using a speckle interferometer \citep{2009AstBu..64..296M} with 550/50, 700/50, and 900/80 nm filters. The average image quality was under 1.2\arcsec. For each series, 2000 frames were acquired with an exposure time of 60~ms. Calibration of the image scale and field orientation was performed following the procedure described in \cite{2020AJ....159..266M}. The measurement of positional parameters and magnitude differences is described in \cite{2005A&A...431..587P}. A comparison of the autocorrelation function images (Figure \ref{fig:saibta}) demonstrates a higher angular resolution attained by the larger mirror of BTA-6 in comparison to SAI-2.5, as the characteristic size of the resulting peaks is proportional to $\lambda/D$, where $\lambda$ is the effective wavelength and $D$ is the telescope aperture.  

\section{Mass estimation}
\label{masses}
\subsection{Mass - luminosity relation}
Binary stars with carefully measured orbital motion allow mass estimates with an accuracy of a few percent \citep{2010A&ARv..18...67T}. However, the considered systems have completed only a small fraction of their orbit since the first resolved observation, and we have to rely on a less direct method to provide a tentative mass estimate. The luminosity of main-sequence stars correlates with mass through the mass-luminosity relation (MLR) \citep{2024PARep...2...41E}. For red dwarfs at visible and infrared wavelengths, the MLR can be described as a continuous and monotonic function which uniquely relates the mass of a star ($m$) to its absolute magnitude ($M$) in a given passband ($p$), such that $M=f_{p}(m)$ and $m=f^{-1}_{p}(M)$. Thus, knowing parallax and the apparent magnitudes of the components, we may constrain their masses.

For red dwarf binary systems with measured dynamical masses, \cite{2016AJ....152..141B} found the intrinsic scatter of the MLR in the visible $V$ band is twice as large as that in the near-infrared $K$ passband, making the latter a preferable choice for photometric mass inference. The discrepancy was attributed to age, metallicity, and magnetic differences of the studied stars. Notably, the $K$ filters at various astronomical instruments have slightly different transmission curves. While the differences are almost negligible (Appendix A of \cite{2019ApJ...871...63M}), our further discussion concerns the specific $K_s$ realization implemented in the 2MASS survey.

The 2MASS survey \citep{2006AJ....131.1163S} remains a principal reference for near-infrared photometry, and we take the apparent $K_s^{\rm app}$ magnitudes from the 2MASS All-Sky Catalog of Point Sources \citep{2003yCat.2246....0C}, where all studied sources were assigned a high photometric quality flag.  With the $2.0 \times 2.0$ arcsec pixel size, binary stars with subarcsecond angular separation appear as unresolved sources in 2MASS. Contamination from unrelated background stars may be an issue, particularly in areas of high stellar density near the Galactic plane. We apply no reddening or extinction correction, as it should be negligible for the objects in the solar neighborhood at infrared wavelengths.

We lack relative photometry measurements in the $K_s$ band from speckle observations and measure the magnitude difference of components with the shorter-wavelength $I_c$ filter instead. This passband currently lacks an empirically calibrated MLR of high precision; therefore, we choose to rely on a theoretical isochrone for a 10 Gyr age from \cite{2015A&A...577A..42B}, which supports several commonly used photometric systems. As discussed in Appendix \ref{isochrones}, the isochrone-predicted mass value is within 6\% of the reliable empirical fit in the $K_s$ band from \cite{2019ApJ...871...63M} for the red dwarfs' mass domain (Figure \ref{fig:models}). We expect a similar, if not larger, discrepancy between empirical data and the theoretical isochrone for the $I_c$ band, thus limiting the accuracy of estimated masses. However, the low photometric accuracy of speckle interferometric observations (Section \ref{observations}) is probably a larger factor of uncertainty.
    
\begin{figure}[ht!]
\plotone{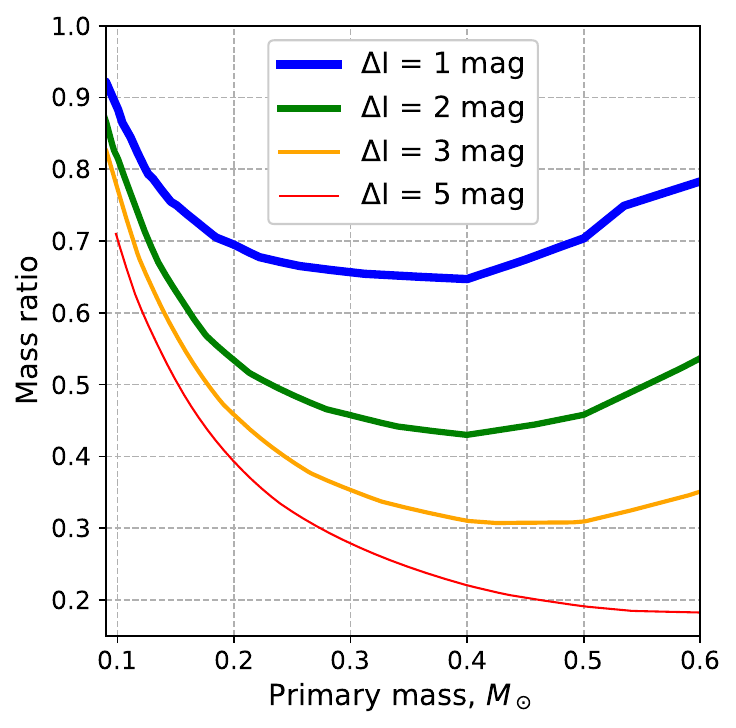}
\caption{The mass ratio as a function of the primary mass for fixed values of the magnitude difference in the $I_c$ band, based on the 10 Gyr isochrone from \cite{2015A&A...577A..42B}. 
\label{fig:q_contrast}}
\end{figure}

Notably, the mass ratio of the components relates to their magnitude difference rather nonlinearly (Figure~\ref{fig:q_contrast}). Thus, while $\Delta I_c = 2$ mag implies $q=0.43$ for a primary star of $m_1=0.4 M_\odot$, for a less massive $m_1=0.1 M_\odot$ primary, the corresponding mass ratio is much larger: $q=0.82$. Since the detectability of a companion in speckle observations depends on the flux contrast, we are less sensitive to companions in the low-mass systems.

\subsection{Photometric estimate}
\label{Photometric}
Three observables are required to constrain the component masses in a binary system: the magnitude difference ($\Delta I_c = I_2-I_1$) measured from speckle observations, the apparent $K_s^{\rm app}$ magnitude of the unresolved source and the trigonometric parallax ($\varpi$); the last two are adopted from the 2MASS and CNS5 catalogs, respectively. These quantities have associated uncertainties, which we consider statistically independent. For each parameter, we generate a set of $10^4$ values assuming a Gaussian distribution. The following steps are then executed independently for each realization:

\begin{enumerate}
    \item Derive the absolute $K_s$ magnitude from the apparent one and the parallax (measured in arcseconds): $K_s=5+5 \log \varpi + K^{\rm app}_s$\\
    This value is to be compared with the estimate deduced from the stellar isochrones in steps 2--6.
    \item Choose a primary mass  ($m_1$) within the $0.08 M_\odot <m_1< 0.7 M_\odot$ range. For the selected value, calculate the $I_c$ and $K_s$ absolute magnitudes using the MLR function for the respective passband:   
    $I_1=f_{\rm I}(m_1)$ ; $K_1=f_{\rm K}(m_1)$
    
    \item Add the magnitude difference from the speckle observations to obtain the secondary $I_c$ magnitude: $I_2=I_1+\Delta I_c$
   
    \item  Derive the secondary mass from $I_2$ using the MLR and calculate the magnitude in the $K_s$ band:\\ 
    $m_2=f^{-1}_{\rm I}(I_2)$ ; $K_2=f_K(m_2)$ 
    
     \item Calculate the combined unresolved magnitude of the primary and secondary components:\\ $K=K_1-2.5 \log (1+10^{0.4(K_1-K_2) })$
   
     \item Find the residual with the value derived in step 1: $\xi=K-K_s$ 
\end{enumerate}
 Steps 2--6 are repeated with different $m_1$ values until the residual falls below the $|\xi|<10^{-5}$ mag threshold. Given that the MLR is a monotonic function, $\xi(m_1)$ changes its sign only once, and we numerically search for the corresponding $m_1$ value with a bisection root-finding method. Once $m_1$ is found, $m_2$ and the magnitudes in different passbands are derived as described above. Thus, we obtain a best-fit solution for a given combination of the $\Delta I_c$, $K^{\rm app}_s$, and $\varpi$. Finally, we calculate the median and the 0.16 and 0.84 quantiles for $10^4$ independent runs, which define the confidence limits.

\subsection{Spectroscopic estimate}
\label{spectroscopic}
For comparison, we also derive the masses based on spectral classification in the literature. The relation between spectral type and mass is taken from Table 16 of \cite{2024ApJS..271...55K}; linear interpolation is adopted. The dependence is notably steep between the M3 and M4 types, where the mass drops by almost a factor of two, from $0.39$ to $0.2 M_\odot$; the estimates are particularly uncertain in this range. The spectroscopic classification comes from different authors and methods and is highly inhomogeneous. For a tentative error estimate, we assign an uncertainty of half a spectral subtype (e.g., M0$\pm0.5$). We assume the reported spectral type belongs to the primary star.  Unlike the photometric approach, the spectroscopic estimate does not require a parallax. A discrepancy between the two methods may indicate an inaccurate parallax value, which is plausible for binary stars, as they often have a large RUWE value or a two-parameter astrometric solution in Gaia DR3.

\section{Unresolved sources}
\label{Unresolved}
It was anticipated that a few sources in our sample would remain unresolved in speckle observations despite their elevated multipeak fraction. Two of our targets have known exoplanets: CNS5 1218 (GJ~179) \citep{2010ApJ...721.1467H} and CNS5 5478 (GJ 849) \citep{2006PASP..118.1685B}. It was hard to expect a detection of an unknown stellar companion to such well-studied objects previously probed with adaptive optics observations at the VLT \citep{2005ApJ...625.1004M} and Keck observatories \citep{2014ApJ...781...28M}. In fact, Proxima~Cen, the closest star to the Sun, has $f_{\rm MP}=11$\% and is absent from our sample solely because of its southern declination. In Appendix \ref{allsky}, we discuss a relative excess of sources with elevated multipeak fraction among nearby stars, which may be a bias in the Gaia DR3 data.

\begin{deluxetable*}{lcccccccc}
\tablecaption{Detection limits for unresolved stars \label{tab:unresolved}}
\tablewidth{0pt}
\tablehead{
 \colhead{CNS5} & \colhead{Date} & \colhead{Band} & \colhead{Seeing} & \colhead{$\varepsilon_{0.1}$} &\colhead{$\varepsilon_{0.2}$}&\colhead{$\varepsilon_{0.4}$}&\colhead{$\varepsilon_{0.6}$}&\colhead{$\varepsilon_{1.0}$} \\
\colhead{} & \colhead{(yr)} & \colhead{} & \colhead{(arcsec)} & \colhead{(mag)} & \colhead{(mag)} & \colhead{(mag)} & \colhead{(mag)}  & \colhead{(mag)}
}
\decimalcolnumbers
\startdata
1218&2025.841&$I_c$&0.78&2.25&3.86&4.85&4.92&5.86\\
1776&2025.841&$I_c$&0.87&1.33&3.69&5.03&5.50&6.32\\
1776&2026.350&$I_c$&0.71&--&2.70&4.22&4.35&6.81\\
3021&2025.929&$I_c$&1.19&--&3.34&5.06&5.07&5.67\\
3021&2026.350&$I_c$&0.82&0.73&3.86&4.42&4.89&5.29\\
4593&2025.846&$I_c$&0.90&--&2.41&4.50&4.80&5.18\\
5362&2025.570&$I_c$&0.95&2.25&4.35&5.84&6.23&7.37\\
5478&2025.573&$880$&0.75&2.63&	4.71&	5.15&	6.56&	7.86\\
5844&2025.570&$I_c$&1.08&2.31&4.66&5.59&6.01&7.09\\
\enddata
\tablecomments{Columns: (1) CNS5 designation; (2) Observational epoch; (3) Passband; (4) Seeing; (5--9) Detection limits at angular separation of 0.1\arcsec, 0.2\arcsec, 0.4\arcsec, 0.6\arcsec and 1.0\arcsec. Detection limits for resolved stars are available at \href{https://doi.org/10.5281/zenodo.23616140}{Zenodo}.}
\end{deluxetable*}

Overall, seven target sources remained unresolved; the obtained detection limits are provided in Table \ref{tab:unresolved}. Six of them have multipeak fraction values within the ${6\%\leq f_{\rm MP}\leq8\%}$ range, close to the ${f_{\rm MP}>5}$\% threshold. The exception is CNS5 4593 with ${f_{\rm MP}=35}$\%. This star is projected onto a high stellar density region near the Galactic plane, which likely explains its elevated $f_{\rm MP}$; the RUWE value of 0.833 is compatible with being a single object. Other unresolved sources also have RUWE $<1.4$, except for CNS5 1776, discussed below.

\subsection{CNS5 1776}
RUWE $= 6.819$ suggests a disturbance in the motion of the source photocenter and the existence of a companion responsible for these deviations. The M1.5 spectral type \citep{2013AJ....145..102L} implies a mass of about 0.5$M_\odot$ and an absolute magnitude of $K_s^{\rm abs}=5.75$ \citep{2013ApJS..208....9P}. Allowing for an uncertainty of 0.1 mag and given the apparent magnitude $K^{\rm app}_s=7.03\pm0.03$ from 2MASS, this star should be at $18\pm1$ pc from the Sun, which agrees with the 19 pc distance derived from the $\varpi=52.76\pm0.12$~mas trigonometric parallax. Hence, a probable companion should have a negligible contribution to the total flux. 

The value of RUWE depends on many circumstances. The forward model of \cite{2024A&A...688A...1C} shows that a value of 6.819 can be reached with a mass ratio as low as $q \sim 0.02$, implying that the companion could be an exoplanet of super-Jupiter type or, more probably, a brown dwarf. Such a low-luminosity object would be out of reach for our instrument. Observations at two different epochs revealed no companion (Table~\ref{tab:unresolved}).

\section{Resolved systems} 
\label{Resolved}

\begin{deluxetable*}{lcccccc}
\tablecaption{Observations of resolved target sources \label{tab:resolved}}
\tablewidth{0pt}
\tablehead{
 \colhead{CNS5} & \colhead{Date} &\colhead{Band} & \colhead{Seeing} & \colhead{Separation} &Position angle&Contrast \\
\colhead{} & \colhead{(yr)} & \colhead{}&  \colhead{(arcsec)} & \colhead{(arcsec)} & \colhead{(deg)} &\colhead{(mag)}
}
\decimalcolnumbers
\startdata
142 & 2025.570 & $I_c$ & 0.90 & $0.165 \pm 0.002$ & $43.14^* \pm 0.56$ & $0.23 \pm 0.15$ \\
142 & 2026.504 & $I_c$ & 0.89 & $0.170 \pm 0.002$ & $72.21^* \pm 0.72$ & $0.00 \pm 0.14$ \\
1280 & 2025.843 & $I_c$ & 1.23 & $0.397 \pm 0.003$ & $159.87 \pm 0.36$ & $0.51 \pm 0.04$ \\
1280 & 2026.641 & $I_c$ & 2.15 & $0.403 \pm 0.007$ & $164.98 \pm 0.94$ & $0.63 \pm 0.14$ \\
1692 & 2025.849 & $I_c$ & 0.91 & $0.229 \pm 0.003$ & $31.68^* \pm 0.86$ & $0.32 \pm 0.12$ \\
1692 & 2026.759 & $I_c$ & 1.14 & $0.216 \pm 0.004$ & $38.25^* \pm 1.09$ & $0.48 \pm 0.16$ \\
1816 & 2025.841 & $880$ & 0.73 & $0.607 \pm 0.003$ & $091.55 \pm 0.29$ & $4.82 \pm 0.09$ \\
1816 & 2026.641 & $I_c$ & 1.42 & $0.571 \pm 0.003$ & $089.99 \pm 0.31$ & $4.87 \pm 0.08$ \\
1816 & 2026.641 & $880$ & 1.46 & $0.565 \pm 0.005$ & $090.16 \pm 0.47$ & $4.72 \pm 0.11$ \\
1907 & 2025.841 & $I_c$ & 0.92 & $0.519 \pm 0.001$ & $071.48 \pm 0.14$ & $1.70 \pm 0.01$ \\
1907 & 2026.641 & $I_c$ & 1.36 & $0.545 \pm 0.001$ & $066.53 \pm 0.14$ & $1.70 \pm 0.01$ \\
2081 & 2026.737 & $I_c$ & 1.20 & $0.478 \pm 0.002$ & $035.80 \pm 0.24$ & $1.68 \pm 0.03$ \\
2267 & 2025.841 & $I_c$ & 0.86 & $0.137 \pm 0.001$ & $067.46 \pm 0.28$ & $0.52 \pm 0.03$ \\
2267 & 2026.350 & $I_c$ & 0.74 & $0.083 \pm 0.001$ & $080.49 \pm 0.21$ & $0.51 \pm 0.02$ \\
3180 & 2025.929 & $I_c$ & 1.33 & $0.397 \pm 0.001$ & $163.66 \pm 0.14$ & $0.43 \pm 0.01$ \\
3180 & 2026.350 & $I_c$ & 1.15 & $0.391 \pm 0.001$ & $161.23 \pm 0.14$ & $0.38 \pm 0.01$ \\
3470 & 2025.575 & $I_c$ & 0.61 & $0.334 \pm 0.004$ & $022.79 \pm 0.62$ & $3.86 \pm 0.07$ \\
3470 & 2026.350 & $I_c$ & 0.93 & $0.276 \pm 0.002$ & $025.79 \pm 0.45$ & $3.72 \pm 0.04$ \\
4695 & 2025.528 & $I_c$ & 0.63 & $0.597 \pm 0.002$ & $017.57 \pm 0.17$ & $2.62 \pm 0.06$ \\
4695 & 2026.506 & $I_c$ & 1.22 & $0.569 \pm 0.005$ & $016.77 \pm 0.52$ & $2.53 \pm 1.32$ \\
4695&2026.810&700&1.36&$0.551\pm0.001$&$016.83\pm0.02$&$3.50\pm0.03$\\
4695&2026.810&900&1.30&$0.552\pm0.001$&$016.48\pm0.02$&$3.00\pm0.02$\\
4709 & 2025.528 & $I_c$ & 0.55 & $0.798 \pm 0.002$ & $302.46 \pm 0.14$ & $3.16 \pm 0.02$ \\
4709 & 2025.570 & $I_c$ & 0.87 & $0.790 \pm 0.002$ & $303.18 \pm 0.16$ & $3.11 \pm 0.03$ \\
4709 & 2026.350 & $I_c$ & 0.68 & $0.794 \pm 0.002$ & $305.63 \pm 0.15$ & $3.15 \pm 0.02$ \\
4709&2026.810&700&1.28&$0.782\pm0.001$&$307.52\pm0.02$&$3.73\pm0.04$\\
4709&2026.810&900&1.09&$0.783\pm0.001$&$307.70\pm0.02$&$2.77\pm0.02$\\
\enddata
\tablecomments{Columns: (1) CNS5 designation; (2) Observational epoch; (3) Passband; (4)~Seeing; (5--7) Binarity parameters and their nominal uncertainties. For CNS5~142 and CNS5 1692, the position angle is determined with a 180$^\circ$ ambiguity. Observations were obtained with the SAI-2.5 telescope (Section \ref{sai25}), apart from measurements with the 700 and 900~nm filters made with BTA-6 (Section~\ref{bta}).} 
\end{deluxetable*}

We resolved eleven target sources as binary stars with an angular separation of ${0.08\arcsec \lesssim \rho \lesssim 0.8\arcsec}$, implying projected distances of ${2 \lesssim s \lesssim 16}$~au between components. Each newly resolved system was observed at least at two different epochs. We provide the binarity parameters in Table \ref{tab:resolved} and the autocorrelation function images in Figure \ref{fig:iconostasis}. In all cases, the observed changes in the components' relative positions are consistent with their orbital motion. Sources corresponding to the detected systems consistently show signs of binarity in Gaia DR3: $f_{\rm MP}\geq 10\%$ and RUWE~$\geq1.685$. All three objects with two-parameter astrometric solutions were resolved.

We found no evidence in the literature that any considered system belongs to a young stellar population or, on the contrary, represents an old, low-metallicity generation of our Galaxy. Owing to their long evolutionary timescales, the luminosity of red dwarfs barely depends on their age. Considering that our targets are typical field objects of unknown age, we provide tentative mass estimates using the photometric and spectroscopic data in Table \ref{tab:masses}. The obtained masses of the primary and secondary components span the entire ${0.075 M_\odot \lesssim m \lesssim 0.61 M_\odot}$ range occupied by M dwarfs \citep{2024ARA&A..62..593H}. While the mean estimates from the two methods may vary, they reasonably agree after the uncertainties are taken into account. Next, we provide more information on specific systems.

\begin{deluxetable*}{lccccccc}
\tablecaption{Tentative mass estimates and projected separation for resolved binaries\label{tab:masses}}
\tablewidth{0pt}
\tablehead{
 \colhead{CNS5} &\colhead{WDS}&Sp&$m_{\rm sp}$& \colhead{$m_1$} & \colhead{$m_2$} & \colhead{$q$} & \colhead{s} \\
&& &\colhead{($M_\odot$)} & \colhead{($M_\odot$)} &\colhead{($M_\odot$)} && \colhead{au}
}
\decimalcolnumbers
\startdata
142&00308+7743&M6$^1$&$0.106^{+0.004}_{-0.004}$&$0.095^{+0.003}_{-0.002}$&$0.093^{+0.002}_{-0.003}$&$0.975^{+0.017}_{-0.027}$&2.2\\
1280&05056+4648&M8$^2$&$0.091^{+0.004}_{-0.004}$&$0.097^{+0.002}_{-0.001}$&$0.091^{+0.002}_{-0.002}$&$0.934^{+0.031}_{-0.027}$&7.0\\
1692&06483+0326&M6$^3$&$0.106^{+0.004}_{-0.004}$&$0.108^{+0.008}_{-0.006}$&$0.100^{+0.006}_{-0.004}$&$0.933^{+0.042}_{-0.047}$&5.5\\
1816&07195+3250&M0$^{4,6}$&$0.59^{+0.01}_{-0.03}$&$0.601^{+0.004}_{-0.003}$&$0.113^{+0.008}_{-0.006}$&$0.189^{+0.013}_{-0.011}$&10.7\\
1907&07422+5004&M2.5$^5$&$0.42^{+0.03}_{-0.03}$&$0.412^{+0.008}_{-0.009}$&$0.201^{+0.017}_{-0.016}$&$0.487^{+0.051}_{-0.044}$&10.9\\
2081&08259+6902&M6$^1$&$0.106^{+0.004}_{-0.004}$&$0.114^{+0.002}_{-0.002}$&$0.088^{+0.001}_{-0.001}$&$0.773^{+0.026}_{-0.023}$&5.9\\
2267&09088+6636&M3.5$^6$&$0.30^{+0.10}_{-0.10}$&$0.239^{+0.011}_{-0.011}$&$0.194^{+0.012}_{-0.012}$&$0.808^{+0.089}_{-0.076}$&2.2\\
3180&12533--0520&M3$^4$&$0.39^{+0.03}_{-0.09}$&$0.270^{+0.064}_{-0.044}$&$0.226^{+0.054}_{-0.036}$&$0.845^{+0.086}_{-0.083}$&8.0\\
3470&14021+2103 &M4$^7$&$0.20^{+0.09}_{-0.09}$&$0.321^{+0.004}_{-0.004}$&$0.096^{+0.003}_{-0.003}$&$0.299^{+0.011}_{-0.011}$&5.6\\
4695&18534+0251&M3$^6$&$0.39^{+0.03}_{-0.09}$&$0.332^{+0.006}_{-0.005}$&$0.123^{+0.008}_{-0.007}$&$0.370^{+0.028}_{-0.026}$&11.6\\
4709&18564+4623&M4.1$^8$&$0.19^{+0.08}_{-0.03}$&$0.244^{+0.002}_{-0.003}$&$0.096^{+0.002}_{-0.003}$&$0.392^{+0.015}_{-0.015}$&15.5\\
\enddata
\tablecomments{Columns: (1) CNS5 designation; (2) Provisional WDS designation; (3) Spectral type with reference: 1 -- \cite{2021PASJ...73..154K}, 2 -- \cite{2024ApJS..271...55K}, 3 -- \cite{2021AA...649A.147S}, 4 -- \cite{2015AA...577A.128A}, 5 -- \cite{2013AJ....145..102L}, 6 -- \cite{2022ApJ...927..122H}, 7 -- \cite{2025RNAAS...9..125L}, 8 -- \cite{2015ApJS..220...16T}; \\ (4) Mass estimate derived from the spectral type (Section \ref{spectroscopic}); (5--7) Mass estimates for both components and mass ratio based on photometric data (\ref{Photometric}); (8) Projected separation.} 
\end{deluxetable*}

\subsection{CNS5 142}
This is probably the nearest newly resolved system in our sample, with a trigonometric parallax of ${\varpi= 76.4\pm4}$~mas \citep{2014ApJ...784..156D}, corresponding to a distance of 13~pc. The source has a two-parameter astrometric solution without reported parallax in Gaia DR3, which itself strongly suggests binarity. \cite{2024ApJS..271...55K} accordingly labeled this star a hidden binary. We resolved CNS5~142 in two observations made a year apart in July of 2025 and 2026. The angular separation barely increased from 0.16\arcsec \  to 0.17\arcsec, while the position angle changed by~$29^\circ$, which is consistent with an orbital motion. The magnitude difference is compatible with being a twin binary. 

 The trigonometric parallax value is challenged by the classification of \cite{2021PASJ...73..154K}, who reported a spectral type of M$6 \pm 0.4$. According to \cite{2013ApJS..208....9P}, an M6 type red dwarf should have an absolute magnitude of $K_s=9.2$. While for a single star this implies a $12 \pm 1$ pc distance, for a twin binary, which is the case here, the system should be at a larger distance of $17 \pm 2$ pc. The photometrically derived primary mass matches the spectroscopic value when $\varpi=64$ mas is adopted (15.6 pc distance).

\subsection{CNS5 1280}

This is the faintest ($G^{\rm app}=15.90$) and least massive primary star in our sample ($m_1 \sim 0.1 M_\odot$), with a spectral type of M8 according to \cite{2024ApJS..271...55K}, who suggested binarity based on its RUWE value (4.823) and position in the color-magnitude diagram. The source is located in a dense region with a Galactic latitude of 3.5$^\circ$, which makes it susceptible to photometric contamination due to crowding. The position angle changed by 5$^\circ$ between November 2025 and August 2026 at an angular separation of $\rho\sim0.4$\arcsec.

\subsection{CNS5 1692}
\label{CNS5_1692}
This star stands out with a large parallax uncertainty due to a two-parameter solution in Gaia DR3. With $\varpi=40.2\pm3.2$ mas \citep{2014ApJ...784..156D}, it lies almost $3\sigma$ below the adopted threshold of $\varpi>50$~mas and is likely beyond 20~pc from the Sun.  Its spectral classification is not directly found in the literature. However, \cite{2021AA...649A.147S} derived an effective temperature of $T= 2831$ K, which almost coincides with the expected value for an M6 dwarf  \citep{2013ApJS..208....9P}. Given that a similar object, 2MASS J02081366+4949023, with $T=2886$ K was classified as M5.5 \citep{2005PASP..117..676R}, we assign an M6 type to CNS5 1692. Both photometric and spectroscopic methods agree on a $0.1 M_\odot$ primary mass estimate. 

\subsection{CNS5 1816}

The binarity was first inferred by \cite{2006ApJ...649..436E}, who explained the observed radial velocity variation by the perturbation from an unknown stellar companion in a long-period orbit. A non-detection in April~2013 was reported by \cite{2022AJ....163..200S} with a limiting contrast of $\Delta I = 2.24$ mag at 0.5\arcsec. In our observations, we achieved better sensitivity and found a companion at $\rho\sim 0.6$\arcsec \ with $\Delta I \sim 4.9$~mag, the largest flux contrast in our sample. With $G^{\rm app}=9.36$~mag, CNS5 1816 is the brightest resolved source in our sample, enabling observations with a narrow-band 880 nm filter to improve the accuracy of the binarity parameters. Between November 2025 and August 2026, the angular separation decreased from 0.61 to 0.57\arcsec,  while the position angle changed by only $1.5^{\circ}$. Considering a proper motion of $|\mu| \sim 0.54$~arcsec~$\cdot$~yr$^{-1}$ in Gaia DR3, a random background star would show a larger movement.

\subsection{CNS5 1907}
\label{CNS5_1907}

This source is marked as a candidate binary by \cite{2025A&A...693A.228C} due to its elevated RUWE and multipeak fraction in Gaia DR3. A star is reported as unresolved by \cite{2020AJ....159..139L}; however, signs of a companion can be seen in Robo-AO observations from January 2020, available at the ExoFOP archive. This database also contains the speckle interferometric observations made in December 2020 at the 8 m Gemini telescope with Alopeke, in which a companion is detected. The reported magnitude contrasts are 2.25 mag and 1.64~mag with the 562~nm and 832~nm filters, respectively. The latter value is close to our $\Delta I_c \sim 1.7$ mag measurement. Since 2020, the angular separation has progressively increased from $\rho=0.37$\arcsec \ to 0.55\arcsec, while the position angle has changed from $\theta=115^\circ$ to $71^\circ$ and $67^\circ$ in November 2025 and August 2026, respectively.

\subsection{CNS5 2081}
\label{CNS5_2081}
Being a known visual binary WDS 08259+6902, this source initially was not part of our survey. However, \cite{2024AJ....167...56C} listed this object as unresolved in their Table 5. Given that CNS5 2081 generally meets our criteria as a nearby source with a large multipeak fraction and unconfirmed binarity (Section \ref{sample}), we undertake an additional observation for clarification. At least four successful detections are reported in the literature. 

\cite{2022AJ....163..200S} first resolved this star in October 2012, measuring ${\rho=0.40\pm0.06\arcsec}$, $\theta=223.0 \pm	8.6^{\circ}$ and $\Delta I= 2.3\pm 0.1$~mag using the 1.5 m telescope at Palomar observatory equipped with the Robo-AO system. Later, \cite{2026AJ....172...40C} reported a detection in April 2017, with $\rho=0.29\arcsec$, $\theta=256^\circ$ and a  magnitude difference of 1.62 mag with a 700 nm filter.

This was followed by an Alopeke speckle interferometric observation in February 2021 at the 8 m Gemini North telescope, available in the ExoFOP archive. It features the smallest angular separation reported for this system, $\rho \sim 0.21\arcsec$. The position angle was originally measured with a $180^\circ$ ambiguity; after taking into account the other observations, we adopt a value of $\theta=342^\circ$. The photometric contrast between components is estimated to be 1.88 mag in the 832 nm filter.

In February 2022 a companion was found at $\rho=0.24\arcsec$ and $\theta=359.0\pm0.5^\circ$ with Keck/NIRC2 adaptive optics imaging \citep{2023AJ....165..162U}. Based on magnitude differences of $\Delta K \sim 1.1$ and $\Delta L \sim 1.0$, the component masses are estimated to be $m_1=0.124$ and $m_2=0.086 M_\odot$, yielding a mass ratio of $q\sim0.7$, which is reasonably close to our $q\sim 0.77$ estimate (Table~\ref{tab:masses}). Finally, in September 2026 we detected a companion at $\rho=0.48\arcsec$, $\theta=36^\circ$ with $\Delta I_c=1.7$ mag. The latter value is consistent with the Alopeke measurement in a similar passband. Overall, CNS5 2081 has completed nearly half of its orbit over the 14 years since the discovery.   

\subsection{ CNS5 2267}

This source has the largest RUWE value in the sample (38.98) and is therefore classified as a candidate binary star by \cite{2025A&A...693A.228C}. It has the smallest angular separation in our sample: $\rho \sim 0.14\arcsec$ \ in November 2025 and $\rho \sim 0.08\arcsec$ \ in May 2026. The position angle changed by 13$^\circ$ over 6 months between the two epochs.

\subsection{CNS5 3180}
With a two-parameter astrometric solution and a 100\% multipeak fraction in Gaia DR3, this source exhibits a full set of binarity indicators. The parallax of $49.5\pm8.9$ mas from ground-based observations \citep{2016AJ....151..160F} is rather imprecise, and an M3 spectral type \citep{2015AA...577A.128A} matches the photometrically derived mass if the binary star is placed at a distance of 28 pc from the Sun.

\subsection{CNS5 3470}

This source has the lowest RUWE (1.685) among the resolved sample objects. Observations in July 2025 and May~2026 show a $3^\circ$ change in the position angle and a decrease in angular separation from 0.33\arcsec to 0.28\arcsec. The modest proper motion of $\mu=0.03$ arcsec$\cdot$yr$^{-1}$ according to Gaia~DR3 leaves a small probability of optical projection instead of genuine binarity for this pair.

\subsection{CNS5 4695}
The source is designated a binary candidate by \cite{2025A&A...693A.228C} due to its large RUWE and multipeak fraction. Three observations obtained between July 2025 and October 2026 show that the angular separation steadily decreased from $\rho=0.59\arcsec$ to 0.55\arcsec, while the position angle remained almost unchanged. Given the proper motion of 0.3 arcsec$\cdot$yr$^{-1}$ from Gaia~DR3, a hypothetical background star would presumably exhibit a larger shift than observed. The binarity of CNS5 4695 is confirmed by observations with BTA-6 (Figure~\ref{fig:saibta}).

\subsection{CNS5 4709}
This is a candidate binary due to its multipeak fraction of 44\% in Gaia DR3 \citep{2025A&A...693A.228C}. This is the widest system in our sample with an angular separation of about 0.8\arcsec. Similarly to CNS5 4695, the proper motion of 0.3 arcsec$\cdot$yr$^{-1}$ eliminates the possibility of optical projection.
The position angle changed by around $5^\circ$ between July 2025 and October 2026, with the latter observation made with BTA-6 (Figure~\ref{fig:saibta}).

\section{Summary and prospects}
\label{summary}

We select 18 sources (Table \ref{tab:targets}) from the Fifth Catalogue of Nearby Stars (CNS5) with reported trigonometric parallaxes possibly exceeding 50 mas, a multipeak fraction value (\text{\it ipd\_frac\_multi\_peak}) in Gaia DR3 larger than 5\% and as yet unconfirmed binarity. For all targets, we conduct speckle interferometric observations with the 2.5~m telescope of SAI MSU observatory (Section~\ref{sai25}). Eleven sources are successfully resolved with angular separations in the range of ${0.08\arcsec \lesssim \rho \lesssim 0.8\arcsec}$ (Table~\ref{tab:resolved}). The tentatively estimated companion masses are within the range of ${0.09 M_\odot \lesssim m_2 \lesssim 0.23 M_\odot}$ (Table \ref{tab:masses}). Both the primary and secondary components are M dwarfs.

For two targets, CNS5 4695 and CNS5 4709, the binarity is confirmed by the speckle interferometric observations made with the 6 m telescope of the SAO RAS observatory (Section \ref{bta}). For CNS5 1907 and CNS5 2081, we found prior high-angular-resolution observations in the ExoFOP archive, including with the 8 m Gemini telescope. All newly resolved systems are promising for precise orbit determination and dynamical mass estimation in the future; their detection improves the completeness of the solar neighborhood census. The discoveries include five potential targets (CNS5~1816, CNS5~1907, CNS5~2267, CNS5~3470, and CNS5~4695) for the Habitable Worlds Observatory \citep{2025PASP..137j4402T}.

Following an analysis of stars that remained unresolved in Section \ref{Unresolved}, we find a possible bias toward elevated multipeak fraction for nearby stars in Gaia DR3, discussed separately in Section \ref{allsky} of the Appendix. 

Six years have passed since the Gaia Early Data Release 3 \citep{2021A&A...649A...1G} went public in December~2020. To a large extent it has become the main reference catalog for studies of stars in the solar neighborhood and beyond \citep{2026PhR..1150....1P}. Today, the major expectations are associated with the upcoming Gaia DR4 release, which will hopefully provide independent confirmation of the discovered binaries. It will introduce 22.8 million solutions for resolved systems, a type of dataset unavailable previously. The new astrometric solutions will take mutual orbital motion into account and improve the reliability of parallaxes, thus bringing the completeness of volume-limited samples to a new level. Meanwhile, we anticipate that even the relatively small samples of objects observed using the speckle interferometric technique will help interpret and navigate the enormous amount of Gaia data lying ahead.

 \section*{Data availability}

Table data, contrast curve and autocorrelation function plots for observations are available at 
\href{https://doi.org/10.5281/zenodo.23616140}{Zenodo}.

\begin{acknowledgments}
Observations at SAI-2.5 were carried out by Polina Budnikova, Ivan Pavlovsky, Nataliia Maslennikova, and Andrei Tatarnikov. The authors thank the engineering and scientific staff of the SAI MSU and SAO RAS observatories, and Andrei Tokovinin for valuable discussions. 

Observations with the SAO RAS telescopes are supported by the Ministry of Science and Higher Education of the Russian Federation. The renovation of telescope equipment is currently provided within the national project "Science and universities". 

The study was conducted under the state assignment of Lomonosov Moscow State University. 

This research has made use of the Exoplanet Follow-up Observation Program (ExoFOP; DOI: 10.26134/ExoFOP5) website, which is operated by the California Institute of Technology, under contract with the National Aeronautics and Space Administration under the Exoplanet Exploration Program.
\end{acknowledgments}

%

\software{Astrophysics Data System\footnote{https://ui.adsabs.harvard.edu/}}, GaiaUnlimited \citep{2024A&A...688A...1C}, SIMBAD \citep{2000A&AS..143....9W}, TOPCAT \citep{2005ASPC..347...29T}, VizieR \citep{2000A&AS..143...23O}


\appendix

\section{Low-RUWE sources with elevated multipeak fraction}
\label{allsky}

During our small survey of nearby stars, we found no companions to six Gaia DR3 sources with a multipeak fraction (\text{\it ipd\_frac\_multi\_peak}, Section~\ref{Multipeak}) in the range of ${6\% \leq f_{\rm MP} \leq 8} \%$, although similar stars in the Pleiades with $f_{\rm MP}>4$\% were successfully resolved using the same instrument and observational technique \citep{2025AJ....169..145C}. The single nature of these objects is further supported by their low RUWE values. Below, we undertake a short study to address this discrepancy.

Here we explore the frequency of sources with low RUWE and an elevated multipeak fraction for the all-sky sample. Initially, we query the Gaia DR3 main source catalog with two conditions imposed on the parallax and the apparent magnitude: $\varpi>20$ mas and $G^{\rm app}<16$ mag. This yields a total of 33170 entries. 
The absolute magnitude is then calculated for each source as $G^{\rm abs}=5+5 \log \varpi + G^{\rm app}$, where $\varpi$ is measured in arcsec.  

\begin{figure}[ht!]
\plotone{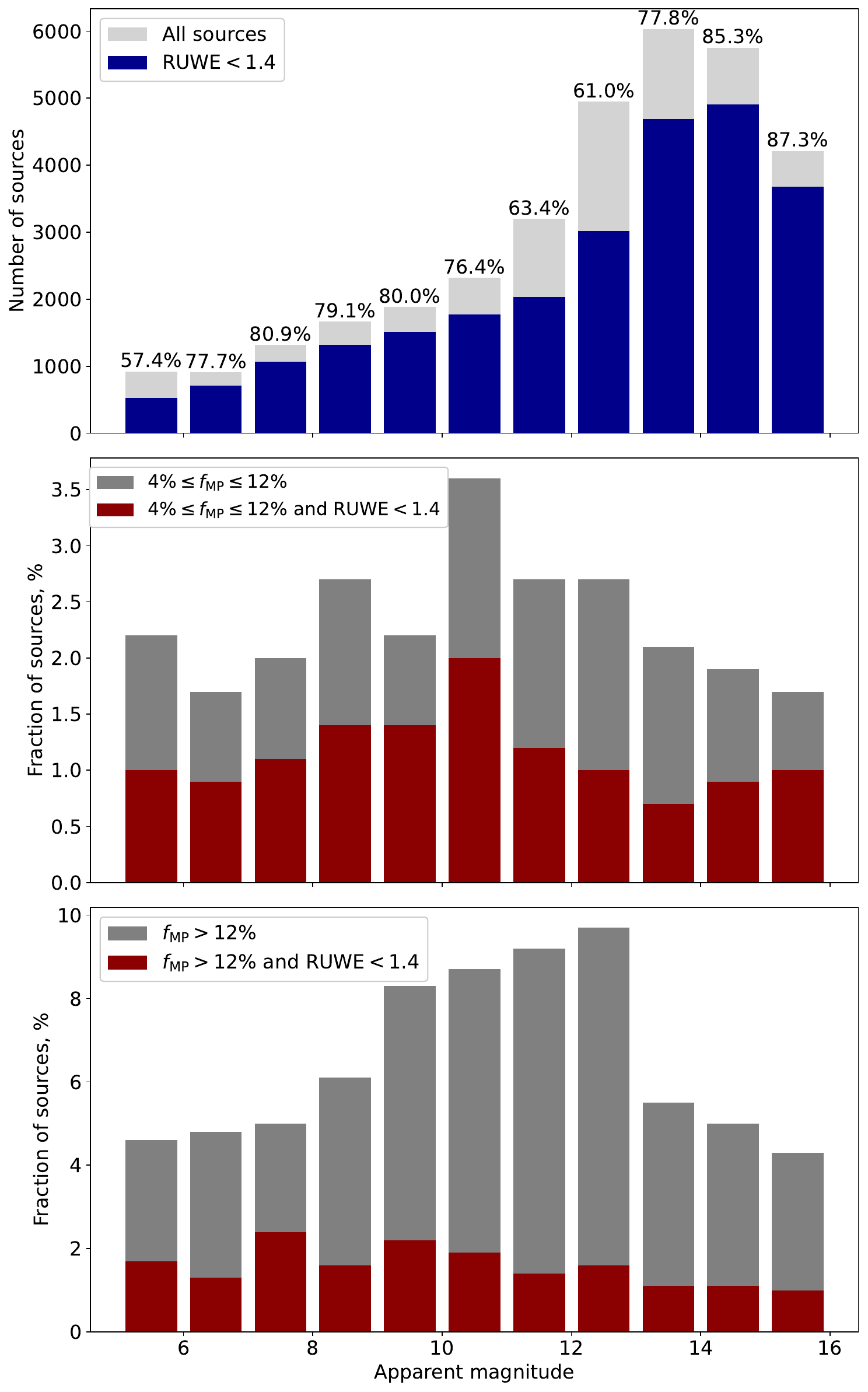}
\caption{The frequency of sources with low RUWE and an elevated multipeak fraction as a function of apparent magnitude. The leftmost bin includes all entries with $G^{\rm app}<6$~mag. In the upper panel, the numbers above the bins indicate the proportion of low-RUWE solutions. In the middle and bottom panels, the relative proportion of sources with the selected $f_{\rm MP}$ and RUWE cuts is shown for each bin.
\label{fig:app_mag}}
\end{figure}

The threshold of RUWE $<1.4$ is often accepted for a reliable astrometric solution in Gaia DR3 that is expected for a single star \citep{2021ApJ...907L..33S}; it is met by a total of 25225 sources, or 76\% of the whole 50~pc sample. However, this fraction is distributed unevenly across apparent magnitude bins (Figure \ref{fig:app_mag}). The low-RUWE solutions appear less frequently for relatively bright stars (57.4\% of sources with $G^{\rm app}<6$~mag) and in the $11<G^{\rm app}<13$ mag range (62.0\%). The highest proportion (86.1\%) is found at the faint end of our sample for sources with $14<G^{\rm app}<16$~mag. 

\begin{figure}[ht!]
\plotone{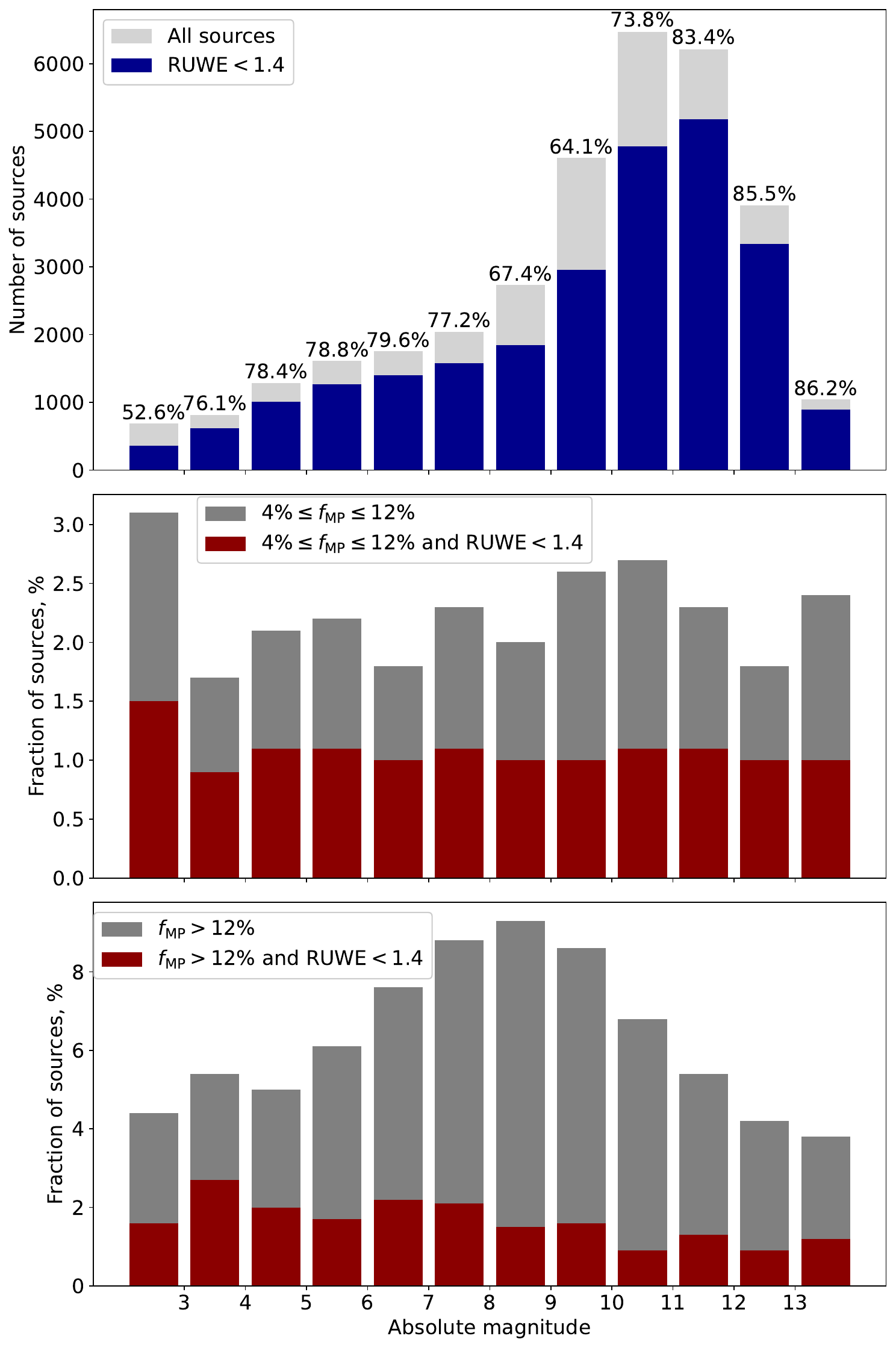}
\caption{The frequency of sources with low RUWE and an elevated multipeak fraction as a function of absolute magnitude. The leftmost and rightmost bins include all entries with $G^{\rm abs}<3.0$ mag and $G^{\rm abs}>13.0$ mag, respectively.
\label{fig:abs_mag}}
\end{figure}

\begin{figure}[ht!]
\plotone{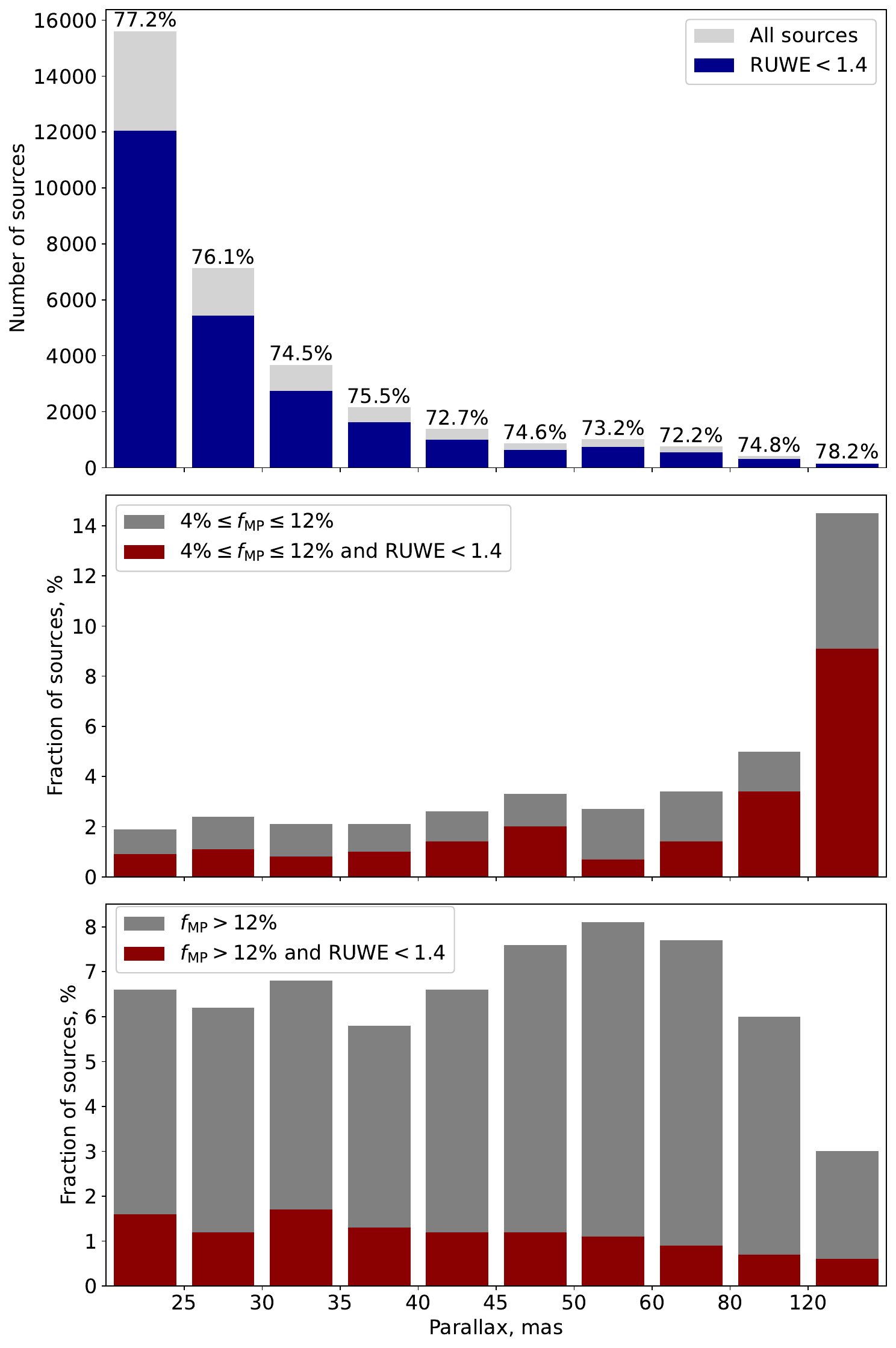}
\caption{The frequency of sources with low RUWE and an elevated multipeak fraction as a function of parallax. The rightmost bin contains all entries with $\varpi>120$ mas.
\label{fig:plx}}
\end{figure}

\begin{figure}[ht!]
\plotone{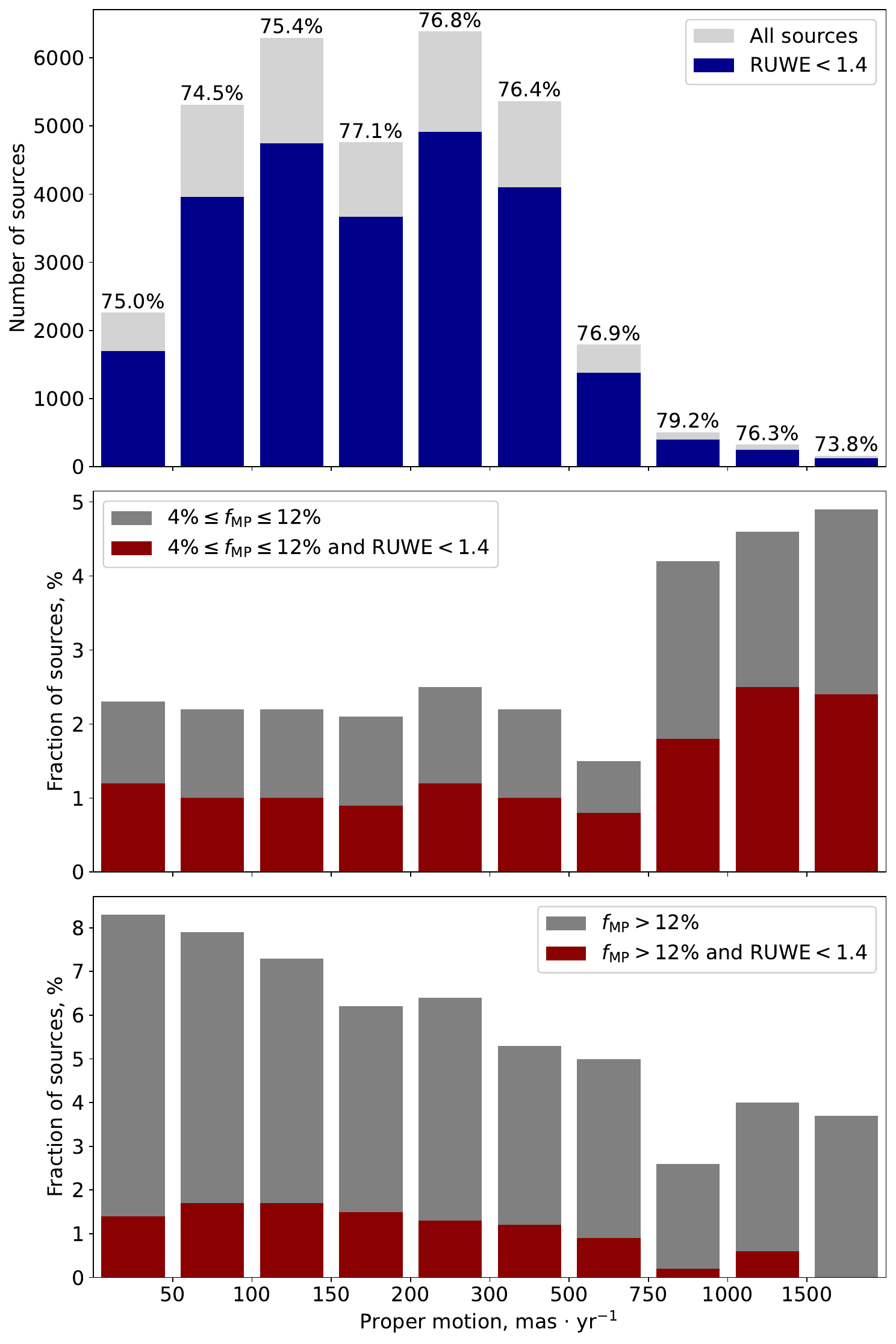}
\caption{The frequency of sources with low RUWE and an elevated multipeak fraction as a function of proper motion. The rightmost bin includes all stars with $\mu>1500$ mas$\cdot$yr$^{-1}$.
\label{fig:pm}}
\end{figure}

The same trend persists for absolute magnitudes (Figure~\ref{fig:abs_mag}). The fraction of solutions with low RUWE increases from 52.6\% for $G^{\rm abs}<3$ mag stars to 84.4\% for ${G^{\rm abs}>11}$~mag sources. Considering that most stars in the solar neighborhood belong to the main sequence, this probably reflects an established fact that stellar multiplicity fraction grows with mass \citep{2023ASPC..534..275O}. At the same time, there is little variation with parallax or proper motion of a source (Figures \ref{fig:plx} and~\ref{fig:pm}).

Similarly to a high RUWE, an elevated multipeak fraction is a useful binarity indicator (Section \ref{Multipeak}). Here we pay most attention to the moderately elevated range of~$4\% \leq f_{\rm MP} \leq 12$\%. Overall, there are 758 solutions meeting this condition, 2.3\% of the total number. The proportion of such solutions relative to the total number of sources is plotted in the middle panels of Figures \ref{fig:app_mag}--\ref{fig:pm}. Unlike for RUWE, the dependence is more prominent for parallax and proper motion, rather than for magnitudes. While there are only 2.2\% of solutions with elevated multipeak fraction among stars with parallaxes of $20<\varpi<80$~mas, this proportion increases to 7.7\% for the nearby stars with $\varpi>80$ mas. A similar trend is found for proper motion: the proportion of solutions with elevated $f_{\rm MP}$ increases from 2.2\% to 4.4\% when the proper motion exceeds the $\nu=0.75$ arcsec $\cdot$ yr$^{-1}$ value.

\begin{deluxetable}{cccc}
\tablecaption{Proportion of sources with elevated multipeak fraction ($4\% \leq f_{\rm MP} \leq 12\%$) relative to the total number within the selected parallax and proper motion cutouts.\label{tab:plx_pm}}
\tablewidth{0pt}
\tablehead{
\colhead{Parameter} &
\colhead{$20<\varpi<80$ mas} &
\colhead{$\varpi>80$ mas} 
}
\startdata
$\nu<750$ mas $\cdot$ yr$^{-1}$& 697/31886 (2.2\%)&  17/289 (5.9\%)   \\
$\nu>750$ mas $\cdot$ yr$^{-1}$ & 16/703 (2.3\%) &  28/292 (9.6\%)
\enddata
\end{deluxetable}

As proper motions and parallaxes of sources are interrelated, we attempt to disentangle their influence on the multipeak fraction. Considering Table \ref{tab:plx_pm}, we notice the proportion of sources with elevated $f_{\rm MP}$ remains low for relatively distant stars regardless of their proper motion, but consistently increases for larger parallaxes. It seems the observed divergence is determined by parallax rather than proper motion, and the nearest stars with $\varpi \gtrsim 80$~mas are more likely to have a moderately elevated $4\%\lesssim f_{\rm MP}\lesssim 12\%$ value in the Gaia DR3 catalog.


Depending on parallax, the same angular separation corresponds to a different orbit size, where the multiplicity properties are inherently different. Considering a mean separation of 40 au for solar-type stars \citep{2023ASPC..534..275O}, genuine binaries may become too wide to trigger an elevated multipeak fraction in Gaia data. Thus, a 40 au orbit at 10 pc distance corresponds to $4\arcsec$.  Indeed, a deficit of solutions with $f_{\rm MP}>12$\% is clearly seen for nearby stars (lower panel of Figure \ref{fig:plx}).

A completely opposite behavior is noticed for the range of $4\%\lesssim f_{\rm MP}\lesssim 12\%$, with a relative excess of such solutions among the nearby sources with ${\varpi>80}$~mas. Simultaneously, another multiplicity indicator, RUWE, does not show a clear dependence on parallax. We remind that we were unable to detect binarity for any of six sources with ${6\%\leq f_{\rm MP}\leq8\%}$ using high angular resolution observations. Therefore, we regard this feature as an observational bias and artifact, rather than an indication of different multiplicity properties.
 
\section{Reference stars}
\label{app:reference}

\begin{deluxetable}{lcccc}
\tablecaption{Detection limits for reference single stars \label{tab:reference}}
\tablewidth{0pt}
\tablehead{
 \colhead{Gaia DR3} & \colhead{Date} & \colhead{Band} & \colhead{0.2\arcsec} &{1.0\arcsec} \\
\colhead{} & \colhead{(yr)} & \colhead{} & \colhead{(mag)} & \colhead{(mag)}
}
\decimalcolnumbers
\startdata
1068811893035042304 & 2025.841 & $I_c$ & 4.41 & 6.89 \\
1097259454301733760 & 2026.737 &  $I_c$  & 4.19 & 5.99 \\
1097349098859057536 & 2026.737 &  $I_c$  & 3.81 & 4.96 \\
1102964098587217664 & 2025.841 & $I_c$ & 4.27 & 7.17 \\
1247279294695975168 & 2026.350 & $I_c$ & 4.25 & 6.46 \\
1926457140524915328 & 2025.570 & $I_c$ & 4.56 & 7.56 \\
1952841438164300160 & 2025.570 & $I_c$ & 4.65 & 7.07 \\
207033916344955904 & 2025.843 & $I_c$ & 3.05 & 4.95 \\
2107258809671762560 & 2025.570 & $I_c$ & 3.68 & 6.38 \\
2627145187596080512 & 2025.573 & $880$ & 3.86 & 7.68 \\
3127358278853055744 & 2025.849 &  $I_c$  & 3.27 & 5.32 \\
3288067846166646272 & 2025.841 &  $I_c$  & 4.08 & 7.04 \\
3584228857953127040 & 2025.929 &  $I_c$  & 4.45 & 7.46 \\
3678364649772785792 & 2025.929 &  $I_c$  & 4.54 & 7.36 \\
4103719342977438080 & 2025.846 &  $I_c$  & 3.27 & 5.28 \\
4279260914134496768 & 2025.528 &  $I_c$  & 4.47 & 7.79 \\
564190305616664192 & 2025.570 &  $I_c$  & 4.19 & 5.78 \\
890234402772937472 & 2026.641 &  $I_c$  & 4.75 & 7.29 \\
\enddata
\tablecomments{Columns: (1) Gaia DR3 designation; (2) Observational epoch; (3) Passband; (4--5) Detection limits at 0.2\arcsec and 1.0\arcsec. Values at other $\rho$ are available at \href{https://doi.org/10.5281/zenodo.23616140}{Zenodo}.}
\end{deluxetable}

Reference single stars required for calibration purposes (Section \ref{observations}) were selected among Gaia DR3 sources with RUWE $<1.25$ and a null multipeak fraction. We report the obtained detection limits for these sources in Table~\ref{tab:reference}. Only one epoch with the largest contrast achieved at $\rho=0.2\arcsec$ is selected. One of the reference objects, 2MASS~J07433735+5004098, a source with ${G=9.45}$~mag and $\varpi=6.948\pm0.016$ mas according to Gaia DR3, was accidentally resolved as a double star. Its binarity parameters are provided in Table~\ref{tab:double_star}. The detected companion may be a mere optical projection.

\begin{deluxetable*}{lcc}
\tablecaption{Serendipitously detected double star\label{tab:double_star}}
\tablewidth{0pt}
\tablehead{\colhead{Parameter} & \colhead{Epoch 1} & \colhead{Epoch 2}
}
\startdata
2MASS&\multicolumn{2}{c}{J07433735+5004098}\\
WDS (provisional)&\multicolumn{2}{c}{07436+5004}\\
Gaia DR3& \multicolumn{2}{c}{982043177014563584}\\
Date (yr)&2025.841&2026.641\\
Seeing (arcsec)&0.88&1.30\\
Separation (arcsec)&0.922 $\pm$ 0.005&0.923 $\pm$ 0.006\\
Position angle (deg)&184.67 $\pm$ 0.30&184.82 $\pm$ 0.37\\
Contrast, $\Delta I_c$ (mag)&5.99 $\pm$ 0.11&5.75 $\pm$ 0.14\\
\enddata

\end{deluxetable*}

\section{Isochrone validation}
\label{isochrones}

\begin{figure}[ht!]
\plotone{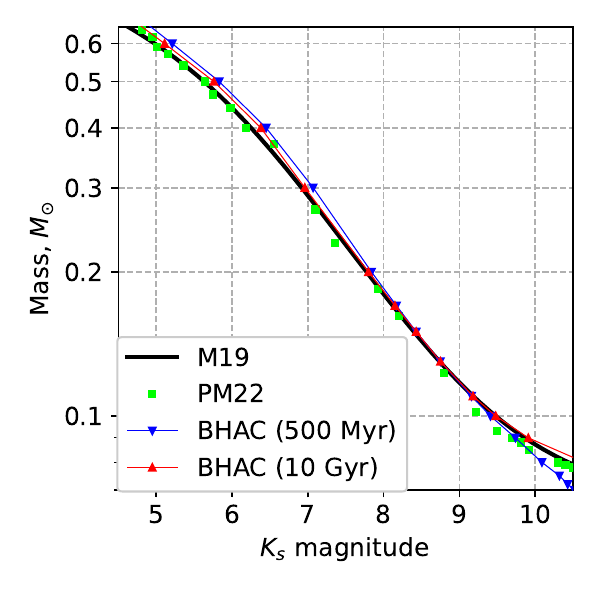}
\caption{Comparison of mass -- luminosity relations in $K_s$ passband from several models. M19: five-degree polynomial fit without metallicity-dependent term from \cite{2019ApJ...871...63M}, PM22: the updated (2022.04.16) sequence of \cite{2013ApJS..208....9P}, BHAC: \cite{2015A&A...577A..42B} isochrones for $t=500$ Myr and 10 Gyr age.
\label{fig:MLR_K}}
\end{figure}

We compare several isochrone models against the mass-luminosity relation of \cite{2019ApJ...871...63M} in the $K_s$ band. The latter is a polynomial function, fitted to dynamically estimated stellar masses of 62 systems with red dwarfs in the solar neighborhood. The reported precision of the predicted masses is around 2\% within the ${4.5<K_s<10.5}$ range of absolute magnitudes.

In Figure \ref{fig:MLR_K} we plot the \cite{2019ApJ...871...63M} polynomial function along with the isochrones of \cite{2015A&A...577A..42B} and an updated (2022.04.16 version) empirical sequence of \cite{2013ApJS..208....9P}. The latter two estimate stellar magnitudes in several bands, including the $K_s$. As the mass-luminosity relation is strictly monotonic, we use linear interpolation to obtain the logarithm of the mass for a star with a given magnitude. Due to slow evolution, ages of red dwarfs are hard to discern \citep{2024ARA&A..62..593H}. 
 
\begin{figure}[ht!]
\plotone{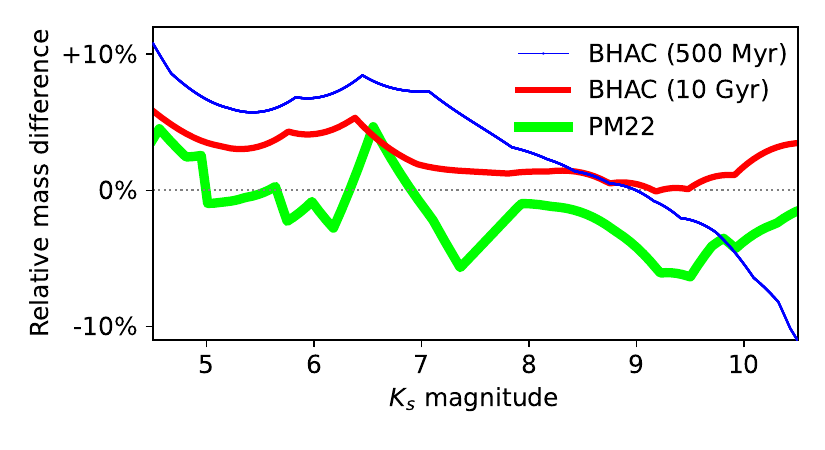}
\caption{Stellar mass ($m$) predicted from several models as a function of $K_s$ magnitude relative to the anchor value ($\hat{m}$) from \cite{2019ApJ...871...63M} fit: $\Delta m / \hat{m}=(m-\hat{m})/\hat{m}$. 
\label{fig:models}}
\end{figure}

\begin{figure}[ht!]
\plotone{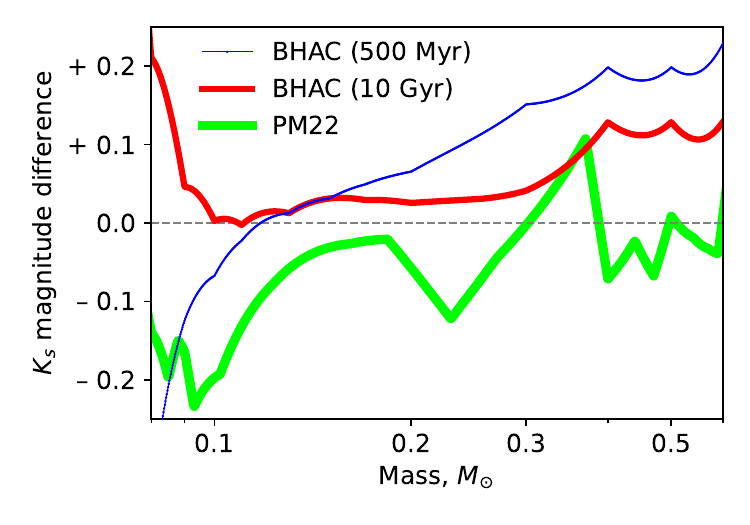}
\caption{The difference of predicted $K_s$ magnitudes as a function of stellar mass relative to the anchor value ($\hat{K_s}$) from \cite{2019ApJ...871...63M} empirical fit: $\Delta K_s = K_s-\hat{K_s}$. 
\label{fig:models2}}
\end{figure}

 The difference in predicted mass depending on $K_s$ magnitude is shown in Figure \ref{fig:models} with the \cite{2019ApJ...871...63M} function as an anchor. The isochrone-predicted masses tend to be overestimated relative to empirical data, although the difference is milder when an age of 10 Gyr is selected. The maximum difference of about 6\% is reached at the bright end, near $K_s=4.5$ mag, while the agreement is within 3\% for the $6.7<K_s<10.2$ mag range. The difference in $K_s$ magnitude as a function of mass is shown in Figure~\ref{fig:models2}.

\bibliography{sample701}{}
\bibliographystyle{aasjournalv7}



\end{document}